\documentclass[twocolumn,aps,prl,superscriptaddress,longbibliography]{revtex4-1}
\usepackage{graphicx}
\usepackage{amsmath}
\usepackage{bm}

\def\bea{\begin{eqnarray}}
\def\eea{\end{eqnarray}}
\def\be{\begin{equation}}
\def\ee{\end{equation}}

\begin{document}

\title{Crossover from the Ultracold to the Quasiclassical Regime in State-Selected Photodissociation}

\author{S. S. Kondov}
\affiliation{Department of Physics, Columbia University, 538 West 120th Street, New York, NY 10027-5255, USA}
\author{C.-H. Lee}
\affiliation{Department of Physics, Columbia University, 538 West 120th Street, New York, NY 10027-5255, USA}
\author{M. McDonald}
\altaffiliation{Present address:  Department of Physics, University of Chicago, 929 East 57th Street GCIS ESB11, Chicago, IL 60637, USA}
\affiliation{Department of Physics, Columbia University, 538 West 120th Street, New York, NY 10027-5255, USA}
\author{B. H. McGuyer}
\altaffiliation{Present address:  Facebook, Inc., 1 Hacker Way, Menlo Park, CA 94025, USA}
\affiliation{Department of Physics, Columbia University, 538 West 120th Street, New York, NY 10027-5255, USA}
\author{I. Majewska}
\affiliation{Quantum Chemistry Laboratory, Department of Chemistry, University of Warsaw, Pasteura 1, 02-093 Warsaw, Poland}
\author{R. Moszynski}
\affiliation{Quantum Chemistry Laboratory, Department of Chemistry, University of Warsaw, Pasteura 1, 02-093 Warsaw, Poland}
\author{T. Zelevinsky}
\email{tanya.zelevinsky@columbia.edu}
\affiliation{Department of Physics, Columbia University, 538 West 120th Street, New York, NY 10027-5255, USA}

\begin{abstract}     
Processes that break molecular bonds are typically observed with molecules occupying a mixture of quantum states and successfully described with quasiclassical models, while a few studies have explored the distinctly quantum mechanical low-energy regime.  Here we use photodissociation of diatomic strontium molecules to demonstrate the crossover from the ultracold, quantum regime where the photofragment angular distributions strongly depend on the kinetic energy, to the energy-independent quasiclassical regime.  Using time-of-flight velocity map imaging for photodissociation channels with millikelvin reaction barriers, we explore photofragment energies in the 0.1-300 mK range experimentally and up to 3 K theoretically, and discuss the energy scale at which the crossover occurs.  Furthermore, we find that the effects of quantum statistics can unexpectedly persist to high photodissociation energies.


\end{abstract}
\date{\today}
\maketitle

\newcommand{\w}{3.25in}

\newcommand{\Schematic}{
\begin{figure}[h]
\includegraphics*[trim = 0.5in 2.4in 0.5in 0.5in, clip, width=3.375in]{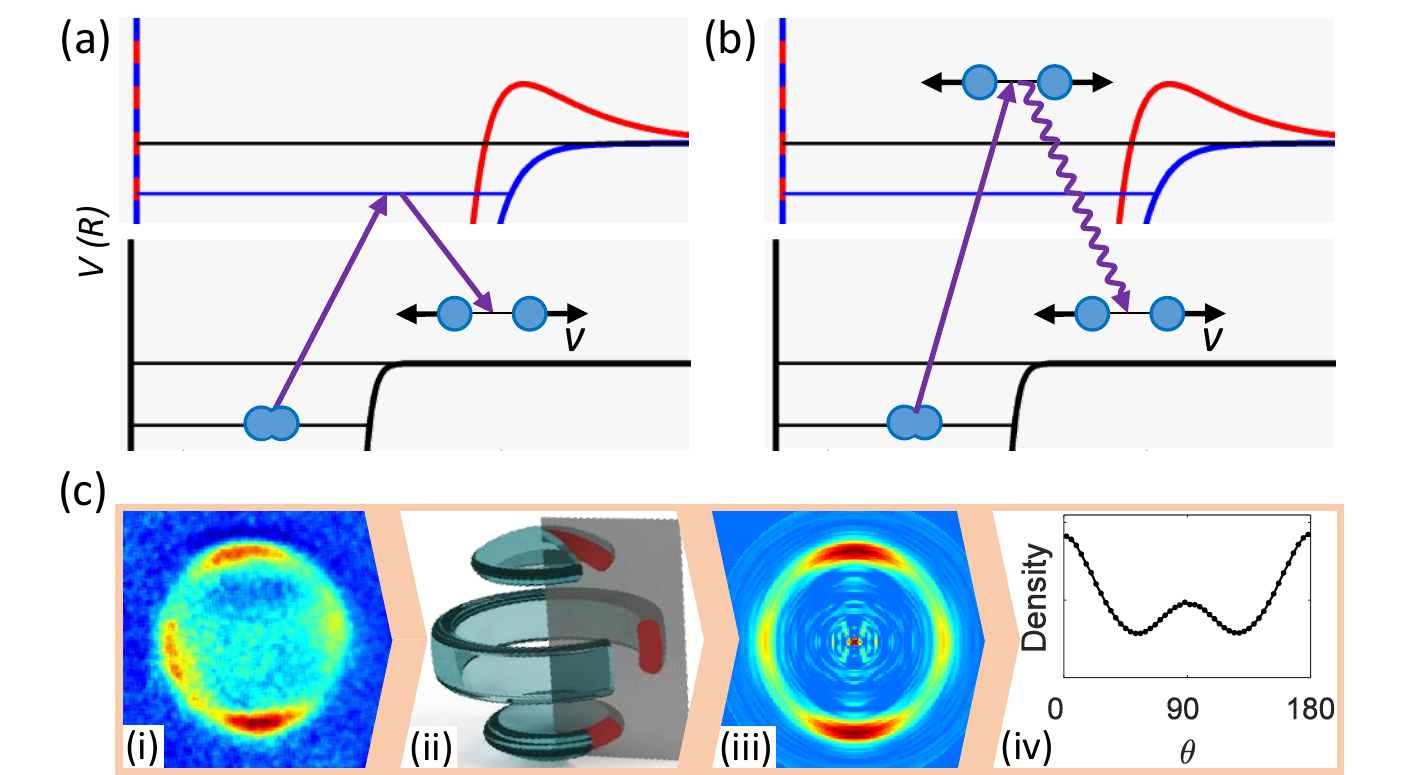}
\caption{(a) Schematic of Sr$_2$ molecule photodissociation to the ground continuum.  The molecules are prepared in a weakly bound vibrational state and fragmented with a resonant two-photon process via the lowest excited electronic state.  The outgoing photofragments have velocity $\nu$ that is determined by the bound-continuum laser frequency, and are detected with absorption imaging.  (b) Photodissociation to the first excited continuum performed with a one-photon process.  This continuum corresponds to a pair of interatomic potentials, $0_u^+$ and $1_u$, where $1_u$ has an electronic barrier.  The excited atomic fragments radiatively decay to the ground state and can be imaged as in (a).  The ground- and excited-state potentials have rotational barriers that are not shown.  (c) Data analysis for azimuthally symmetric photofragment distributions.  A time-of-flight absorption image of photofragments, typically 0.3-0.5 mm in diameter (i), is a line-of-sight integral of the three-dimensional distribution (ii, shown as a surface of constant density).  A cross section of this distribution, shown in red, is retrieved (iii) and radially averaged to yield an angular density profile (iv).}
\label{fig:Schematic}
\end{figure}
}

\newcommand{\PADsExperiment}[1][\w]{
\begin{figure}[h]
\includegraphics*[trim = 1.2in 1.3in 1.2in 0.2in, clip, width=3.375in]{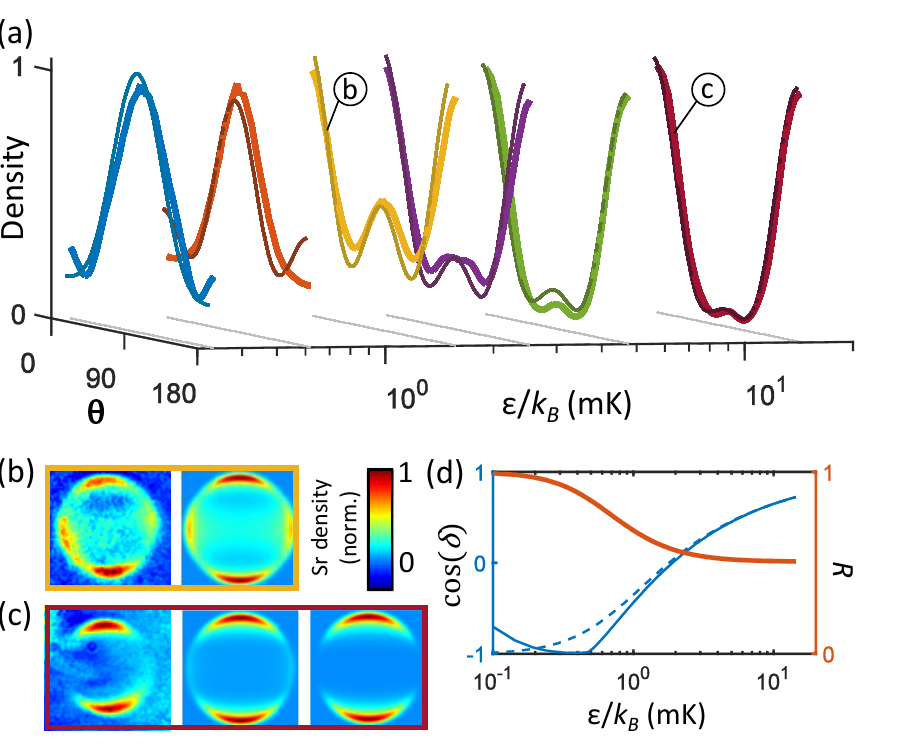}
\caption{(a) Angular density profiles of photofragments as a function of continuum energy, for the $0_u^+(-4,1,0)$ initial state (thick lines:  experiment; thin lines:  \textit{ab initio} theory).  The two labeled curves correspond to (b,c), where curve (c) nearly matches the quasiclassical expectation.  (b) Experimental photofragment absorption image and the corresponding quantum chemistry calculation for the process in (a) at the continuum energy $\varepsilon/k_B=1.6$ mK ($\varepsilon/h=33$ MHz; $h$ is the Planck constant) where quantum mechanical behavior dominates.  (c) Absorption image for $\varepsilon/k_B=14$ mK ($\varepsilon/h=300$ MHz) on the left, with the corresponding quantum chemistry model, as well as the axial recoil limit on the right.  (d) Calculated energy evolution of $R$ (thick red line) and $\cos\delta$ (thin solid line) for the process in (a-c).  The thin dashed line shows the WKB approximation and its agreement with the quantum theory at $\varepsilon/k_B\gtrsim1$ mK.}
\label{fig:PADsExperiment}
\end{figure}
}

\newcommand{\NoQuasiclassical}[1][\w]{
\begin{figure}[h]
\includegraphics*[trim = 2in 2.5in 2.8in 0.5in, clip, width=3.375in]{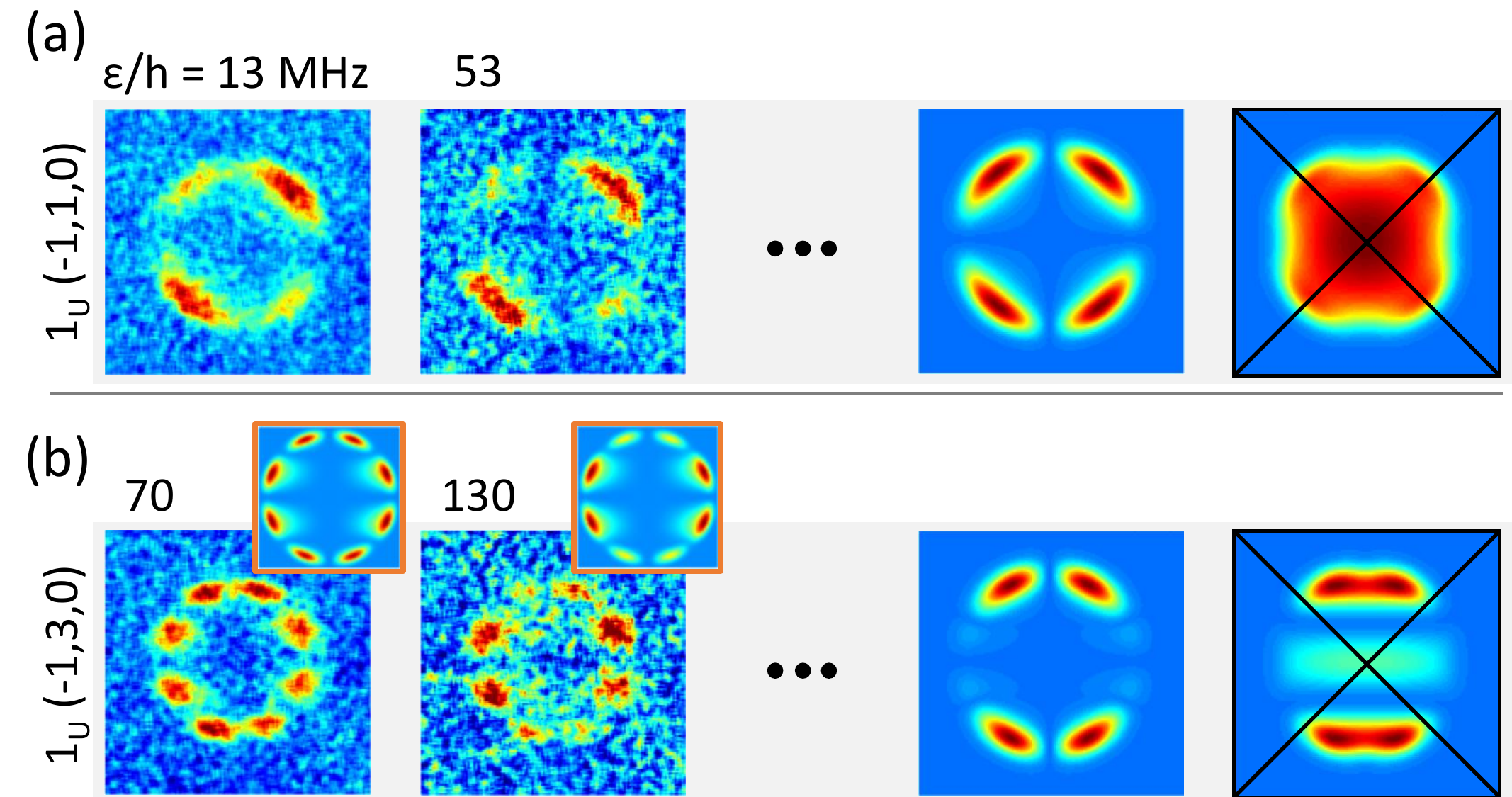}
\caption{Quantum statistics of identical particles prevents agreement with quasiclassical predictions at large photofragment energies.  The cases shown here use $P=1$ polarization.  (a) Photofragment angular distributions for the $1_u(-1,1,0)$ molecules at $\varepsilon/k_B=\{0.63,2.5\}$ mK ($\varepsilon/h=\{13,53\}$ MHz) on the left along with the quantum mechanical and quasiclassical predictions on the right.  This case is energy-independent.  The quasiclassical picture fails to describe the process due to quantum statistics, or the missing $J=1$ partial waves in the ground-state continuum.  (b) Energy-dependent angular distributions for the $1_u(-1,3,0)$ molecules at $\varepsilon/k_B=\{3.4,6.3\}$ mK ($\varepsilon/h=\{70,130\}$ MHz) are shown on the left, where the insets show the corresponding calculations.  The high-energy quantum mechanical and quasiclassical predictions are shown on the right.  While the highest energy regime could not be reached experimentally due to very weak bound-continuum transition strengths and a limited laser intensity, at lower energies the experiment fully agrees with quantum mechanical calculations (shown in the insets).}
\label{fig:NoQuasiclassical}
\end{figure}
}

\newcommand{\UngeradeHighEnergy}[1][\w]{
\begin{figure}[h]
\includegraphics*[trim = 3in 1.7in 3.5in 0in, clip, width=3.375in]{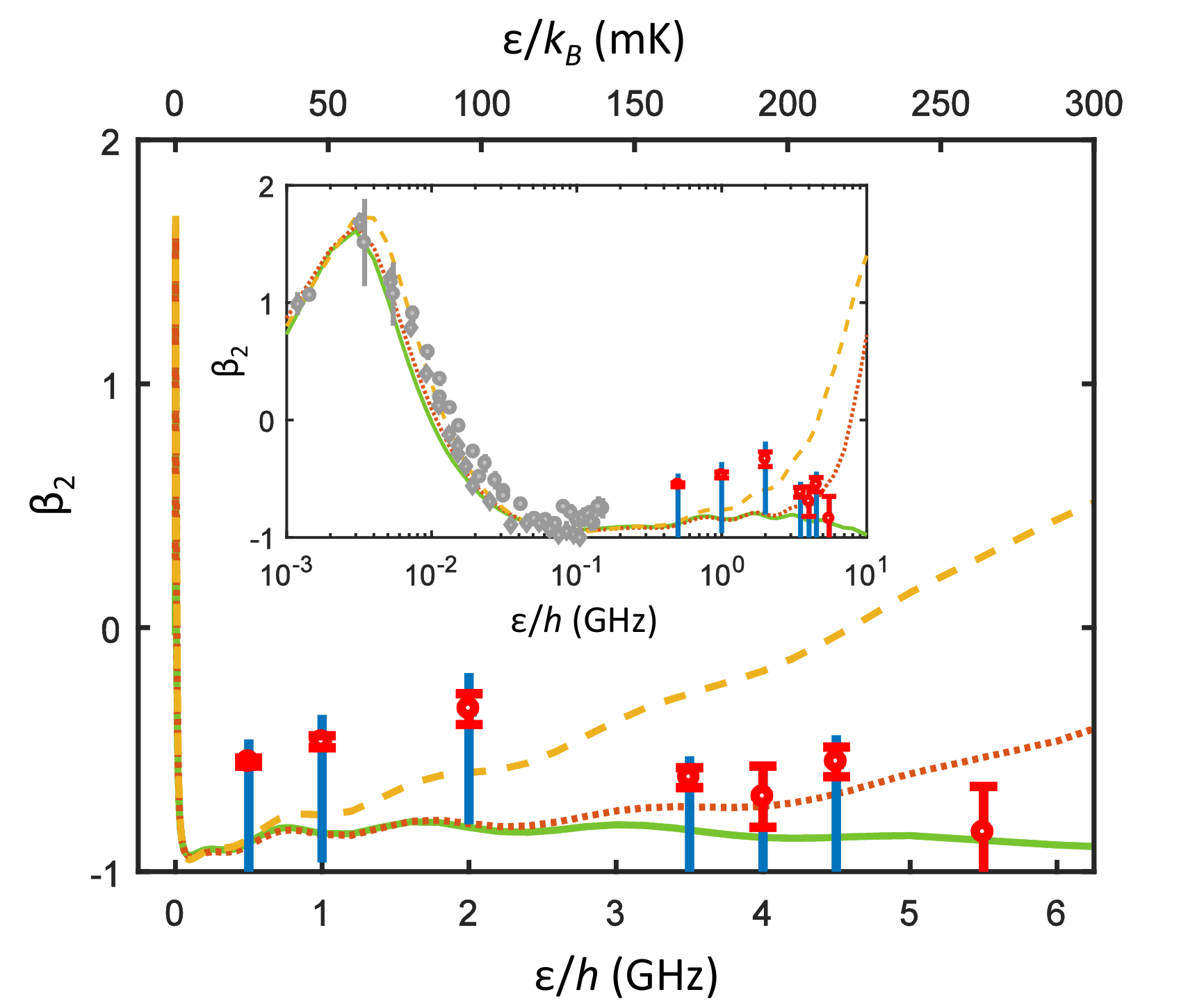}
\caption{Anisotropy parameter $\beta_2$ measured for photodissociation of ground-state $X0_g^+(-1,0,0)$ molecules to the singly-excited $\{0_u^+,1_u\}$ continuum.
The red capped error bars are determined from a bootstrap analysis of up to 1,000 experimental realizations, and the blue capless error bars result from possible contamination by the molecules initially in $J_i=2$.  The dotted red line corresponds to the \textit{ab initio} molecular potential \cite{MoszynskiSkomorowskiJCP12_Sr2Dynamics}, the solid green line to the potential that is optimized to reproduce long-range properties \cite{KillianBorkowskiPRA14_SrPAMassScaling}, and the dashed yellow line to the \textit{ab initio} potential that was manually fitted for a better spectroscopic agreement with weakly bound states \cite{ZelevinskyMcGuyerNJP15_Sr2Spectroscopy}.  Photodissociation of these very weakly bound molecules does not yet reach the expected high-energy limit of $\beta_2=2$ at the experimentally accessible continuum energies up to 260 mK (5.5 GHz).  The inset shows the high-energy data in the context of previous measurements \cite{ZelevinskyMcDonaldNature16_Sr2PD} with improved \textit{ab initio} theory.}
\label{fig:UngeradeHighEnergy}
\end{figure}
}

Recent progress in atomic and molecular physics has resulted in the development of an unprecedented degree of control over molecular degrees of freedom.  This includes targeted preparation of molecular samples in specific internal quantum states as well as in well-defined motional states.  The ability to create molecular samples at a range of cold and ultracold temperatures has enabled detailed studies of few-body chemistry in the regime where quantum mechanical effects, such as resonant scattering and barrier tunneling, define the reaction cross sections.

Approaches to exploring the formation and breaking of molecular bonds across the low energy regime include a variety of techniques such as photoassociation \cite{JonesRMP06}, ultracold atomic collisions \cite{DenschlagWolfScience17_RbStateToStateChemistry,KjaergaardThomasNComm16_UltracoldFermionCollider}, ultracold molecular collisions \cite{YeOspelkausScience10_KRbReactions}, state-selected atomic and molecular beams with relative velocity control \cite{ZarePerreaultScience17_1KelvinMolecularCollisions,MeijerGilijamseScience06_OHXeTunableCollisions,MeerakkerVogelsScience15_NOHeCollisionResonances,NareviciusKleinNPhys17_H2HeScatteringResonances,BethlemPoelPRL18_MolecularSynchrotron,RempeWuScience17_CryofugeColdCollisions}, and molecule collisions with cold trapped ions \cite{SoftleyWillitschPRL08_ColdIonMoleculeCollisions}.  Each method has its benefits such as high tunability of collision energies, the ability to access ultralow energies, or the possibility to handle diverse molecular species.  We have studied molecular photodissociation \cite{SatoCR01_PhotodissociationReview} in the ultracold regime as a path to detailed quantitative understanding of ultracold chemistry phenomena \cite{ZelevinskyMcDonaldNature16_Sr2PD}.  If the molecules are trapped at sub-millikelvin temperatures, optical control of the bond breaking process offers the advantages of reaching the near-threshold regime with extremely high energy resolution and tunability.  The resolution is limited only by the natural lifetime of the photofragments and the initial trap temperature (both typically $<5$ mK, and in this work $\sim1$ $\mu$K due to narrow intercombination transitions), and the tunability only by the laser intensity and field of view.  Furthermore, this reverse-collision technique permits studies of any collisional thresholds that can be reached with one or a few photons, even those with short natural lifetimes that would prevent a successful measurement in a scattering configuration.

Previously we reported photodissociation of ultracold Sr$_2$ molecules in the near-threshold quantum mechanical regime \cite{ZelevinskyMcDonaldNature16_Sr2PD} including interference of photofragment matter waves, sensitivity to reaction barriers, shape resonances, and control of reactions by magnetic fields \cite{ZelevinskyMcDonaldPRL18_PDMagneticField}.  The measured photofragment angular distributions disagreed with the quasiclassical intuition to varying degrees \cite{ZelevinskyMcDonaldNature16_Sr2PD}.  It remained an open question whether a crossover into quasiclassical behavior could be observed and explained from first principles.  Here, we predict and observe the crossover from the quantum mechanical ultracold-chemistry regime to the quasiclassical regime of photodissociation.  We show that the photofragment angular distributions exhibit strong variations with the continuum energy, but stabilize to energy-independent quasiclassical patterns at energies that exceed the reaction barrier height.  We find that photodissociation of very weakly bound molecules can exhibit quantum mechanical behavior to still higher energies.  Our study includes an electronically excited multichannel continuum in addition to the ground-state continuum.  Furthermore, we describe and clarify different levels of approximation that can be applied to predict the photofragment angular distributions \cite{ZareMPC72_PhotoejectionDynamics,BernsteinChoiJCP86_StateSelectedPhotofragmentation,ZareCPL89_PhotofragmentAngularDistributions,SeidemanCPL96_MagneticStateSelectedPDDistributions,ZareBeswickJCP08_PhotofragmentAngularDistrQuantClass,AshfoldWredeJCP02_AxialRecoilBreakdown}.  Finally, we show that the bosonic or fermionic quantum statistics of the photofragments can prevent the photodissociation outcome from reaching the quasiclassical limit even at high energies.

In the experiment, $^{88}$Sr atoms are laser cooled and photoassociated in a one-dimensional optical lattice, yielding $\sim7,000$ Sr$_2$ molecules trapped at a few microkelvin \cite{ZelevinskyReinaudiPRL12_Sr2}.  The lattice trap at the wavelength of $\sim910$ nm has a $\sim30$ $\mu$m radius and a $\sim730$ $\mu$m length.  The molecules predominantly occupy the most weakly bound vibrational level, $v=-1$, in the electronic ground state that correlates to the atomic ${^1S}+{^1S}$ state.  They are distributed between two angular momenta $J_i=\{0,2\}$, either of which can be selected as the starting state for photodissociation, with selectivity of the projection quantum number $M_i$.  Alternatively, weakly bound levels that correlate to the singly excited long-lived ${^1S}+{^3P}_1$ continuum can be populated prior to photodissociation by 689 nm light that co-propagates with the lattice.  The photodissociation light pulses are 10-20 $\mu$s, the photofragment time of flight varies from $\sim800$ $\mu$s near threshold to $\sim20$ $\mu$s at the higher energy range, and the imaging pulse length is 10-20 $\mu$s long.  The absorption imaging beam is resonant with the strong 461 nm Sr transition, nearly co-aligned with the lattice, and expanded to $\sim300$ $\mu$m in order to intercept the outgoing photofragments \cite{ZelevinskyMcDonaldNature16_Sr2PD}.  The (vertical) laboratory quantization axis is set by the lattice polarization, or by a $\sim3$ G magnetic field when required for state selection, while the photodissociation light has polarization that is parallel ($P=0$) or perpendicular ($P=1$) to this axis.  The continuum energy is determined by the frequency of the photodissociation light.  Reaching high energies can be challenging because of the diminishing bound-continuum transition strengths and the rapid expansion of the photofragments.

\Schematic
Figure \ref{fig:Schematic}(a,b) shows two photodissociation processes used in this work.  In case (a), a single molecular quantum state $(v,J_i,M_i)$ of the $0_u^+$ or $1_u$ electronic manifold is resonantly populated and simultaneously photodissociated to the ground continuum $X0_g^+$, while process (b) samples the electronically excited continuum from a single ground molecular quantum state (the excited atomic fragment spontaneously decays prior to imaging).  The upper continuum has contributions from both the barrierless $0_u^+$ potential and the $1_u$ potential with a $\sim1$ mK electronic barrier, where the potentials are labeled by $\Omega_i$, the total atomic angular momentum projection onto the internuclear axis.  Rotational barriers present for all continuum states with angular momentum $J\neq0$ are not shown.  An image of a photofragment angular distribution and the data analysis procedure are illustrated in Fig. \ref{fig:Schematic}(c).  Panels (i-iv) show an azimuthally symmetric time-of-flight (velocity map) image, a three-dimensional distribution that results in this image via line-of-sight integration, the cross section of the distribution obtained via the inverse Abel transform, and the radial average of the cross section showing the measured angular photofragment density.

We have explored experimentally and theoretically the crossover from ultracold to quasiclassical chemistry, and the degree of applicability of a range of approximations.  The quantum mechanical treatment involving
bound and continuum wave functions and using Fermi's golden rule to calculate the photodissociation cross sections is in agreement with data across all sampled energies and for molecules in all initial quantum states, with further improvement possible only by introducing small corrections to the \textit{ab initio} molecular potentials.  It is necessary to use the quantum mechanical treatment to model our observations near threshold.  On the other hand, at high energy the axial recoil limit is reached, where photodissociation
is much faster than molecular rotation and the photofragments emerge along the molecular axis.  The main questions we address are (1) at what energy scale do the angular distributions approach the axial recoil limit; (2) how does this scale depend on the quantum numbers and binding energy of the initial molecule; and (3) how can quantum state selection of the molecules prevent the high-energy axial recoil limit from agreeing with quasiclassical intuition?

In related work, we address the applicability of the WKB approximation and of a semiclassical model (that considers classical rotation of the molecule during photodissociation) \cite{ZelevinskyInPrepQMQCLong18}.  In general we find that (i) near threshold, the quantum mechanical treatment correctly captures the observed photofragment angular distributions and their dependence on the continuum energy; (ii) the axial recoil limit is reached at energies that exceed any electronic and rotational barriers in the continuum; (iii) for very weakly bound molecules, quantum effects can dominate at much higher energies than specified in (ii); and (iv) while the axial recoil approximation is usually equivalent to the ubiquitous quasiclassical model, this is not the case if additional selection rules are imposed by bosonic or fermionic nature of the photofragments.

\PADsExperiment
Figure \ref{fig:PADsExperiment} illustrates the evolution of an angular distribution as a function of the continuum energy $\varepsilon$ for the $0_u^+(v=-4,J_i=1,M_i=0)$ initial state and $P=0$ (as in Fig. \ref{fig:Schematic}(a)), which is a feature of near-threshold photodissociation.  The molecules are photodissociated over $\sim2$ orders of magnitude of continuum energies, with Fig. \ref{fig:PADsExperiment}(a) displaying the angular photofragment densities as a function of $\varepsilon/k_B$ where $k_B$ is the Boltzmann constant.  Quantum chemistry calculations of the expected density curves, based on \textit{ab initio} Sr$_2$ potentials \cite{MoszynskiSkomorowskiJCP12_Sr2Dynamics,KillianBorkowskiPRA14_SrPAMassScaling}, are overlayed with the data.  The measured images for $\varepsilon/k_B=1.6$ and $14$ mK are shown in the leftmost panels of Fig. \ref{fig:PADsExperiment}(b,c), followed by the theoretical images.  Figure \ref{fig:PADsExperiment}(c) also shows an image calculated for the axial recoil limit, which is already approached at 14 mK.  For the case shown, the angular distribution in the axial recoil limit, $I_{\mathrm{AR}}(\theta,\phi)$, agrees with the quasiclassical model:  $I_{\mathrm{AR}}(\theta,\phi)=I_{\mathrm{QC}}(\theta,\phi)$.  The quasiclassical model predicts the angular distribution as a product of the initial molecular orientation probability density $P_i(\theta,\phi)$ that is a function of the polar and azimuthal angles $\{\theta,\phi\}$ referenced to the quantization axis, and the angular probability density of the photodissociation light absorption, $I_{\mathrm{QC}}(\theta,\phi)\propto P_i(\theta,\phi)[1+\beta_2P_2(\cos\theta)]$ \cite{BernsteinChoiJCP86_StateSelectedPhotofragmentation,ZareCPL89_PhotofragmentAngularDistributions}.  Here $P_2$ is the Legendre polynomial and $\beta_2$ is the anisotropy parameter such that $\beta_2=2$ for a parallel photodissociation transition ($\Delta\Omega=0$) resulting in a dipolar photofragment distribution along the quantization axis and $\beta_2=-1$ for a perpendicular transition ($|\Delta\Omega|=1$) with a photofragment distribution transverse to the axis.

For molecules composed of identical constituents such as bosonic $^{88}$Sr atoms, spin statistics imposes selection rules on the allowed angular momenta through the required symmetry under nuclear exchange.  In the electronic ground state and in excited states with $\Omega_i=0$, only even $J$ values are allowed for $^{88}$Sr$_2$.
Since odd $J$ are forbidden in the ground state, this continuum can be represented with only two parameters $\{R,\delta\}$, such that the amplitudes of finding the photofragments in $J=J_i-1$ and $J=J_i+1$ are $\sqrt{R}$ and $\sqrt{1-R}$, respectively, with a phase difference $\delta$.
The energy evolution of the $R$ and $\delta$ parameters is plotted in Fig. \ref{fig:PADsExperiment}(d).  In the axial recoil limit, $R\approx1/2$ and $\cos\delta=1$ \cite{ZelevinskyInPrepQMQCLong18}.  While the angular distributions in Fig. \ref{fig:PADsExperiment}(a-c) and the $\{R,\delta\}$ parameters in Fig. \ref{fig:PADsExperiment}(d) show a strong dependence on $\varepsilon/k_B$ up to $\sim5$ mK, we confirm that at the higher energies they approach stable quasiclassical distributions, with $R$ showing a faster convergence than $\delta$.  The experiments were performed for a range of weakly bound $0_u^+$ and $1_u$ states \cite{ZelevinskyInPrepQMQCLong18}, confirming this convergence as well as the initial-state-dependent evolution for lower energies.  For $1_u$ molecules, due to the perpendicular nature of the electric-dipole transition to the ground-state continuum, at high energies the fragments emerge horizontally (at $90^{\circ}$ to the light polarization) rather than vertically as for $0_u^+$ in Fig. \ref{fig:PADsExperiment}(b,c).

Quantum statistics of the photofragments can affect the reaction outcome.  For example, in photodissociation of state-selected $^{88}$Sr$_2$ molecules to the ground-state continuum some reaction channels (odd $J$) are excluded.  This inherently quantum effect influences the photofragment distributions even at high energies.  We investigate this phenomenon by measuring and calculating the angular distributions for the photodissociation pathway in Fig. \ref{fig:Schematic}(a) with $\Omega_i=1$.  We find that if $\{M_i=0,P=0\}$ are not both true \cite{ZelevinskyInPrepQMQCLong18}, the quantum mechanical angular distributions do not match the quasiclassical predictions in the high-energy limit.

\NoQuasiclassical
Figure \ref{fig:NoQuasiclassical} shows two examples of photodissociation for selected initial states such that the resulting angular distributions do not converge to the quasiclassical picture.  Spin-statistics selection rules lead to energy-independent patterns in one case while the other case is energy-dependent ($P=1$ for both).  In Fig. \ref{fig:NoQuasiclassical}(a), the $1_u(-1,1,0)$ molecules are dissociated at 0.63 and 2.5 mK (13 and 53 MHz) above threshold.  Due to the $\Delta J=\pm1$ and $\Delta M=\pm1$ selection rules, only $J=2$ is allowed in the continuum, and therefore no energy dependence is expected \cite{ZelevinskyInPrepQMQCLong18}.  The data confirms an unchanging angular distribution that matches the quantum mechanical prediction and clearly fails to match the quasiclassical model.  Figure \ref{fig:NoQuasiclassical}(b) illustrates energy-dependent photodissociation of $1_u(-1,3,0)$ molecules.  Here, the near-threshold energy dependence arises from the interference of the $J=2$ and $J=4$ continuum states allowed by optical and spin-statistics selection rules.  The right-hand panels show the calculated quantum mechanical angular distribution in the axial recoil limit that disagrees with the quasiclassical approximation.  To demonstrate agreement with the quantum mechanical model, the left-hand side shows angular distributions at 3.4 and 6.3 mK (70 and 130 MHz) above threshold that match the calculated distributions in the insets.  While the axial recoil regime ($>50$ mK) was not reached in this case due to weak bound-continuum transition strengths and $\sim1$ mW maximum available photodissociation laser power, this limitation is not fundamental and can be overcome with a higher laser intensity.  Note that if optical selection rules (rather than spin-statistics restrictions) allow only a single partial wave $J$ in the continuum, then quantum mechanical and quasiclassical angular distributions strictly agree \cite{ZelevinskyMcDonaldNature16_Sr2PD}.

\UngeradeHighEnergy
A key feature of photodissociation is the ability to select one of many possible continua.  In Fig. \ref{fig:UngeradeHighEnergy} we photodissociate ground-state $X0_g^+(-1,0,0)$ molecules to the singly-excited \textit{ungerade} continuum, and sample energies in the range of 0.07-260 mK (1.5-5,500 MHz).  Here the electronic potential barrier height is only $\sim1$ mK, being proportional to the very small $C_3$ dispersion coefficient that is determined by the inverse of the metastable $^3P_1$ atomic lifetime.  The photofragments have a single angular momentum quantum number $J=1$ but two possible values of $\Omega=\{0,1\}$ that are mixed via nonadiabatic Coriolis coupling, especially at the lower energies \cite{ZelevinskyMcGuyerPRL13_Sr2ZeemanNonadiabatic}.  This mixing (but not interference, unlike for different partial waves $J$ \cite{ZelevinskyInPrepQMQCLong18}) has a strong and nontrivial effect on photofragment angular distributions.  In this case of a spherically symmetrical initial molecular state, the angular distributions can be described as $I_{\mathrm{QC}}(\theta)$ for all continuum energies, but with a varying $\beta_2(\varepsilon)$ that becomes constant at the axial recoil limit.  Figure \ref{fig:UngeradeHighEnergy} shows the plot of $\beta_2(\varepsilon)$ across the wide energy range that is limited only by the $\sim3$ mW photodissociation laser power, where the smaller error bars are limited by the image quality and the larger ones conservatively estimate possible contamination by molecules initially in $J_i=2$.  In the case of mixed $\Omega$ quantum numbers in the continuum, the quasiclassical picture does not predict which $\Omega$ dominates at high energy and whether the observed pattern will tend to a perpendicular ($\Omega=1$) or a parallel ($\Omega=0$) dipole.  For the experiment in Fig. \ref{fig:UngeradeHighEnergy}, the \textit{ab initio} pattern tends to a parallel dipole ($\beta_2=2$) in the axial recoil limit, but for the very weakly bound molecules this regime is expected to be reached only at $>0.5$ K above threshold.  In the energy regime that is currently accessible, the photofragment angular distributions vary steeply with energy in the region of the $\sim1$ mK electronic barrier, then stabilize at $\beta_2\approx-1$.  The energy interval where $\beta_2\approx-1$ is sensitive to long-range molecular potentials, as we have confirmed by adjusting the $C_6$ coefficients.  The measurements in this higher energy regime allow us to distinguish between the \textit{ab initio} \cite{MoszynskiSkomorowskiJCP12_Sr2Dynamics}, long-range \cite{KillianBorkowskiPRA14_SrPAMassScaling}, and fitted \textit{ab initio} \cite{ZelevinskyMcGuyerNJP15_Sr2Spectroscopy} potentials to which the angular distributions are sensitive, as shown in Fig. \ref{fig:UngeradeHighEnergy}.

In conclusion, we have explored experimentally and theoretically how the ultracold, quantum mechanical regime of state-selected photodissociation crosses over into the classical regime at increasing photofragment energies.  The question of applicability of quasiclassical descriptions to photodissociation reactions has lingered in the literature for several decades \cite{HerschbachZarePIEEE63_DiatomicPhotodissociation,ZareMPC72_PhotoejectionDynamics,SeidemanCPL96_MagneticStateSelectedPDDistributions,ZareBeswickJCP08_PhotofragmentAngularDistrQuantClass,ZelevinskyMcDonaldNature16_Sr2PD}, and this work presents a conclusive resolution of this debate.  Additional details are supplied in Ref. \cite{ZelevinskyInPrepQMQCLong18}.  We find that the high-energy axial recoil limit is reached when the continuum energies exceed any electronic and rotational barriers, although quantum effects can dominate to larger energies for very weakly bound molecules.  We study a hierarchy of approximations (WKB, semiclassical, and axial recoil) \cite{ZelevinskyInPrepQMQCLong18}, and experimentally confirm that the commonly used quasiclassical formula for photofragment angular distributions \cite{BernsteinChoiJCP86_StateSelectedPhotofragmentation,ZareCPL89_PhotofragmentAngularDistributions,SeidemanCPL96_MagneticStateSelectedPDDistributions} correctly describes the axial recoil limit for a variety of initial molecular states with different sets of quantum numbers, while in the ultracold regime there is a strong nonclassical variation of the angular distributions with energy.  We demonstrate that the effects of spin statistics for identical photofragments can persist to indefinitely large photodissociation energies and prevent the angular distributions observed in the axial recoil limit from approaching the quasiclassical picture.  Finally, we probe a molecular continuum with a mixture of $\Omega$ quantum numbers in an energy range of over three orders of magnitude and with fine resolution, resolving between the \textit{ab initio} potentials and those that have been adjusted using molecular spectroscopy.  Photodissociation of ultracold molecules with individually isolated quantum states uniquely enables accurate studies of molecular continua, and for relatively simple molecules such as Sr$_2$ the state-of-the-art quantum chemistry theory yields excellent agreement with measurements.  These features enable us to directly observe and accurately model the crossover from ultracold to quasiclassical chemistry.

We acknowledge the ONR Grants No. N00014-17-1-2246 and N00014-16-1-2224, as well as the NSF Grant No. PHY-1349725.  R. M. and I. M. also acknowledge the Polish National Science Center Grant No. 2016/20/W/ST4/00314 and M. M. the NSF IGERT Grant No. DGE-1069240.  We are grateful to C. Liedl and K. H. Leung for their contributions to the experiment.

%

\end{document}